\documentclass{article}
\usepackage{spconf}
\usepackage{graphicx}
\usepackage{amsmath}
\usepackage{amssymb}
\usepackage{multirow}
\usepackage{balance}
\usepackage{xcolor}
\usepackage{url}
\usepackage{verbatim}
\usepackage{enumitem}
\usepackage{pifont}
\usepackage[ruled]{algorithm2e}
\usepackage{fix-cm}

\SetKwInOut{KwInput}{Input}                
\SetKwInOut{KwOutput}{Output}
\SetKwRepeat{Repeat}{repeat\{}{until}%
\DeclareMathOperator*{\argmin}{arg\,min}
\newcommand\tab[1][0.1cm]{\hspace*{#1}}
\newcommand{\size}[2]{{\fontsize{#1}{0}\selectfont#2}}

\title{MONSTR: Model-Oriented Neutron Strain Tomographic Reconstruction}

\name{\em Mohammad Samin Nur Chowdhury$^{1}$, Shimin Tang$^{2}$, Singanallur V. Venkatakrishnan$^{3}$, \\ \em   Hassina Z. Bilheux$^{2}$, Gregery T. Buzzard$^{4}$, and Charles A. Bouman$^{1}$}
\vspace{-1in}
\address{$^{1}$School of Electrical and Computer Engineering, Purdue University, IN 47907, USA\\ 
$^{2}$Neutron Scattering Division, Oak Ridge National Laboratory, TN 37830, USA\\ 
$^{3}$Electrical and Engineering Infrastructure Division, Oak Ridge National Laboratory, TN 37831, USA\\ 
$^{4}$Department of Mathematics, Purdue University, IN 47907, USA \thanks{This manuscript has been authored by UT-Battelle, LLC, under contract DE-AC05-00OR22725 with the US Department of Energy (DOE). The US government retains and the publisher, by accepting the article for publication, acknowledges that the US government retains a nonexclusive, paid-up, irrevocable, worldwide license to publish or reproduce the published form of this manuscript, or allow others to do so, for US government purposes. DOE will provide public access to these results of federally sponsored research in accordance with the DOE Public Access Plan (http://energy.gov/downloads/doe-public-access-plan).
}
}

\begin{document}
\setlength{\abovedisplayskip}{0pt} \setlength{\abovedisplayshortskip}{0pt}
\maketitle

\begin{abstract}
Residual strain, a tensor quantity, is a critical material property that impacts the overall performance of metal parts.
Neutron Bragg edge strain tomography is a technique for imaging residual strain that works by making conventional hyperspectral computed tomography measurements, extracting the average projected strain at each detector pixel, and processing the resulting \textit{strain sinogram} using a reconstruction algorithm.
However, the reconstruction is severely ill-posed as the underlying inverse problem involves inferring a tensor at each voxel from scalar sinogram data.

In this paper, we introduce the model-oriented neutron strain tomographic reconstruction (MONSTR) algorithm that reconstructs the 2D residual strain tensor from the neutron Bragg edge strain measurements.
MONSTR is based on using the multi-agent consensus equilibrium framework for the tensor tomographic reconstruction.
Specifically, we formulate the reconstruction as a consensus solution of a collection of agents representing detector physics, the tomographic reconstruction process, and physics-based constraints from continuum mechanics.
Using simulated data, we demonstrate high-quality reconstruction of the strain tensor even when using very few measurements.
\end{abstract}


\section{Introduction}
\label{sec:introduction}

\begin{figure*}[t!]
\centering
\centerline{\includegraphics[width=0.9\linewidth]{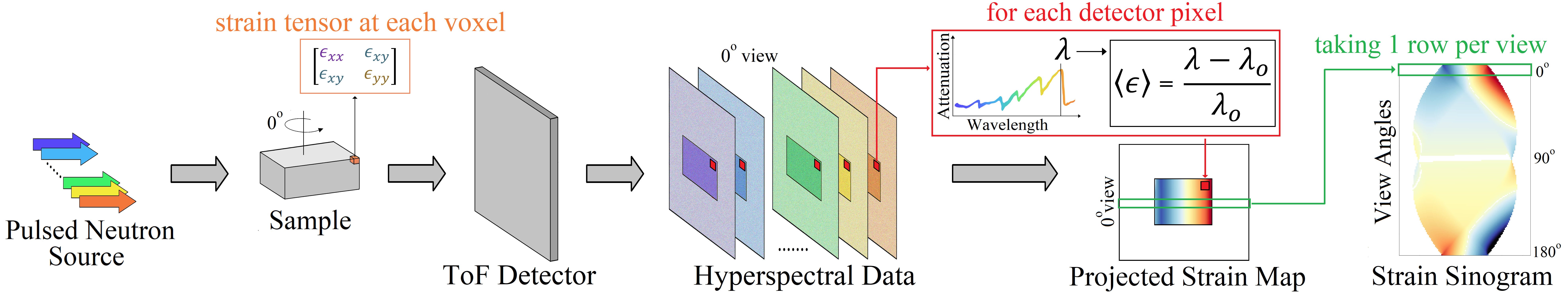}}
\vspace{-0.4cm}
\caption{
Experimental setup for strain sinogram measurement:
Hyperspectral neutron data are collected using a pulsed source and a TOF detector.
For each detector pixel, the Bragg edge location ($\lambda$) is estimated and compared to a strain-free reference to compute a projected strain map.
This process is repeated across multiple views.
For the 2D case, a single detector row is selected, and the projected strain map entries along this row across all views are combined to form a strain sinogram.
}
\vspace{-0.4cm}
\label{fig:strain_sino_gen}
\end{figure*}

Residual strain is the permanent deformation that persists in a material after the removal of external forces, loads, or thermal gradients.
It is a tensor quantity because it has directional components, describing how the material deforms in different directions.
Residual strain can arise from factors such as manufacturing processes (e.g., welding \cite{feng2005welding}, forging \cite{robinson2010forging}, or casting \cite{garza2007casting}), phase transformations during thermal cycles \cite{belyaev2014phase}, and mechanical loading (e.g., cyclic \cite{james2010cycling} or overloading \cite{dalaei2011overloading}).
Although certain residual strains are intentionally induced to enhance material properties, others can lead to undesirable effects such as stress concentrations, crack initiation, or failure under operational conditions.
Therefore, accurate measurement and analysis of residual strain are essential to evaluate a component's strength, durability, and longevity.

Neutron diffraction is a well-established technique for imaging residual strain within manufactured parts \cite{noyan1987diffraction}.
It measures the strain tensor in a small region of the part by placing it in the path of a hyperspectral neutron beam, collecting the diffracted signal using a bank of detectors, and processing the data from each detector bank to obtain a component of the strain tensor. 
In order to image the strain across the entire sample, the measurement is repeated by rastering the sample in the path of the neutron beam, one location at a time. 
Because the time taken to measure the strain tensor from a single location can be long (10s of seconds), the overall scan time can be of the order of days for a large volume.

Neutron Bragg edge strain tomography is a relatively new technique for imaging the residual strain distribution within a sample.
A typical scan works by making conventional hyperspectral computed tomography (CT) measurements, followed by extracting the average projected strain at each detector pixel based on the shift of spectral edges (known as Bragg edges) compared to a strain-free reference, as shown in Fig.~\ref{fig:strain_sino_gen}.
The resulting \textit{strain sinogram} is then used to reconstruct the strain tensor for the entire sample.
The reconstruction is extremely challenging because the underlying inverse problem is severely ill-posed since we have to infer a tensor at each voxel from scalar sinogram measurements. 
Furthermore, the measurement time in a typical hyperspectral neutron CT can be of the order of hours, allowing for the collection of only a sparse number of projections. 

The early approaches for neutron Bragg edge strain tomography made several simplifying assumptions about the underlying strain pattern \cite{gregg2017axisym, kirkwood2015axisym,hendriks2017insitu} that significantly reduced the number of unknowns in the inverse problem, but also limited their applicability.
Some recent efforts have introduced solutions for general 2D residual strain estimation, leveraging reduced-order representation such as a Fourier basis \cite{gregg2018general} or a Gaussian Random process \cite{hendriks2019general} combined with physics-based constraints to reduce the problem space.

In this paper, we present the model-oriented neutron strain tomographic reconstruction (MONSTR) algorithm that performs fast and high-quality reconstruction of the 2D residual strain tensor from a strain sinogram.
MONSTR is the first algorithm to use a multi-agent consensus equilibrium (MACE) framework \cite{bouman2022mbir} for 2D strain tensor tomography, which reduces the inverse problem’s degrees of freedom by enforcing physical constraints through modular, task-specific spatial and sinogram domain agents.
MONSTR offers the following novel contributions:
\begin{itemize}[leftmargin=1em]
\vspace{-0.05cm}
\itemsep-0.4em
    \item Simplifies physical constraint integration by assigning each physical constraint to a dedicated agent, avoiding the complexity of formulating a large joint problem.
    \item Provides flexibility to add or remove agents based on problem-specific constraints.
    \item Enables the use of existing well-developed scalar tomography software (e.g., SVMBIR \cite{svmbir-code}) for tensor tomography.
\vspace{-0.05cm}
\end{itemize}

\vspace{-0.8cm}
\section{Forward Model for neutron Bragg edge strain tomography}
\label{sec: strainsino}

Fig.~\ref{fig:strain_sino_gen} shows a typical setup for neutron Bragg edge strain tomography, which involves placing the sample in the path of a spectral neutron beam and measuring the transmission signal using a time-of-flight (ToF) detector \cite{carminati2020bragg}.
For polycrystalline materials, the ToF transmission spectra contain characteristic signatures with sharp changes known as Bragg edges.
When there is residual strain in the sample, it impacts the location of these Bragg edges.
A standard pre-processing step is to extract the location of a particular Bragg edge for all detector pixels and create a projected strain map based on the relative shifts of the Bragg edge compared to some known strain-free reference.
If  $\lambda_i$ is the Bragg edge location estimated from the $i^{th}$ detector measurement, the corresponding projected strain $\langle\epsilon\rangle_i$ is calculated as:
\begin{equation} \label{eq:lambda}
    \langle\epsilon\rangle_i = \frac{\lambda_i -\lambda_o}{\lambda_o} \ ,
    \vspace{-0.18cm}
\end{equation}
where $\lambda_o$ is the Bragg edge location for a strain-free sample.
This computation is repeated for different orientations/views of the sample. 
For this paper, we consider only a 2D strain tensor using data from a single detector row.
The strain map entries along the selected row from all views are then combined to construct a strain sinogram (see Fig.~\ref{fig:strain_sino_gen}).

We formulate a forward measurement model for this problem using the longitudinal ray transform (LRT) \cite{lionheart2015lrt, polyakova2023lrt} as demonstrated in \cite{gregg2018general}. 
We use the following notation:
\begin{itemize}[leftmargin=1em]
\vspace{-0.05cm}
\itemsep-0.4em
    \item \size{9.3} {$\epsilon = [\epsilon_{xx}, \tab \epsilon_{yy}, \tab \epsilon_{xy}]^\top$ is the strain tensor to be reconstructed, where $\top$ denotes transpose.  This is a vector field with a 3D vector at each voxel location. We use $\epsilon_k$ to denote the component scalar fields, where $k=xx,yy,xy$.}
    \item \size{9.3} {$A$ is tomographic projection operator from volume to sinogram.}
    \item \size{9.3} {$p_k = A\epsilon_k$ where $k=xx,yy,xy$.  We call this a \textit{virtual sinogram} since it is not directly measured.}
    \item \size{9.3} {$p = [p_{xx}, \tab p_{yy}, \tab p_{xy}]^\top$ is the tensor of strain virtual sinograms.  }
\vspace{-0.05cm}
\end{itemize}
We let $L_i$ be the thickness of the sample along the projection for the $i^{th}$ sinogram entry, which is assumed to be known in advance. We let $\theta_i$ be the associated projection angle relative to the $x$-axis.
According to LRT, the $i^{th}$ strain sinogram entry $\langle\epsilon\rangle_i$ is:
\vspace{0.2cm}
\begin{align} \label{eq:forward}
    \nonumber
    \langle\epsilon\rangle_i & = \frac{1}{L_i} \big\{p_{xx,i} \cos^2(\theta_i) +  p_{yy,i} \sin^2(\theta_i) + p_{xy,i} \sin(2\theta_i) \big\} \\
    & = \frac{1}{L_i} w_i p_i \ ,
    \vspace{-0.2cm}
\end{align}
where $w_i = [\cos^2(\theta_i),\tab \sin^2(\theta_i),\tab \sin(2\theta_i)]$ is the weight array for the $i^{th}$ projection.
In summary, from the strain tensor $\epsilon$, we use the projection operator $A$ to obtain the virtual sinograms in $p$, then take the linear combination defined in eq.~\ref{eq:forward} to obtain the corresponding projected strain tensor $\langle\epsilon\rangle$.

\begin{figure}[b!]
\vspace{-0.2in}
\centering
\centerline{\includegraphics[width=\linewidth]{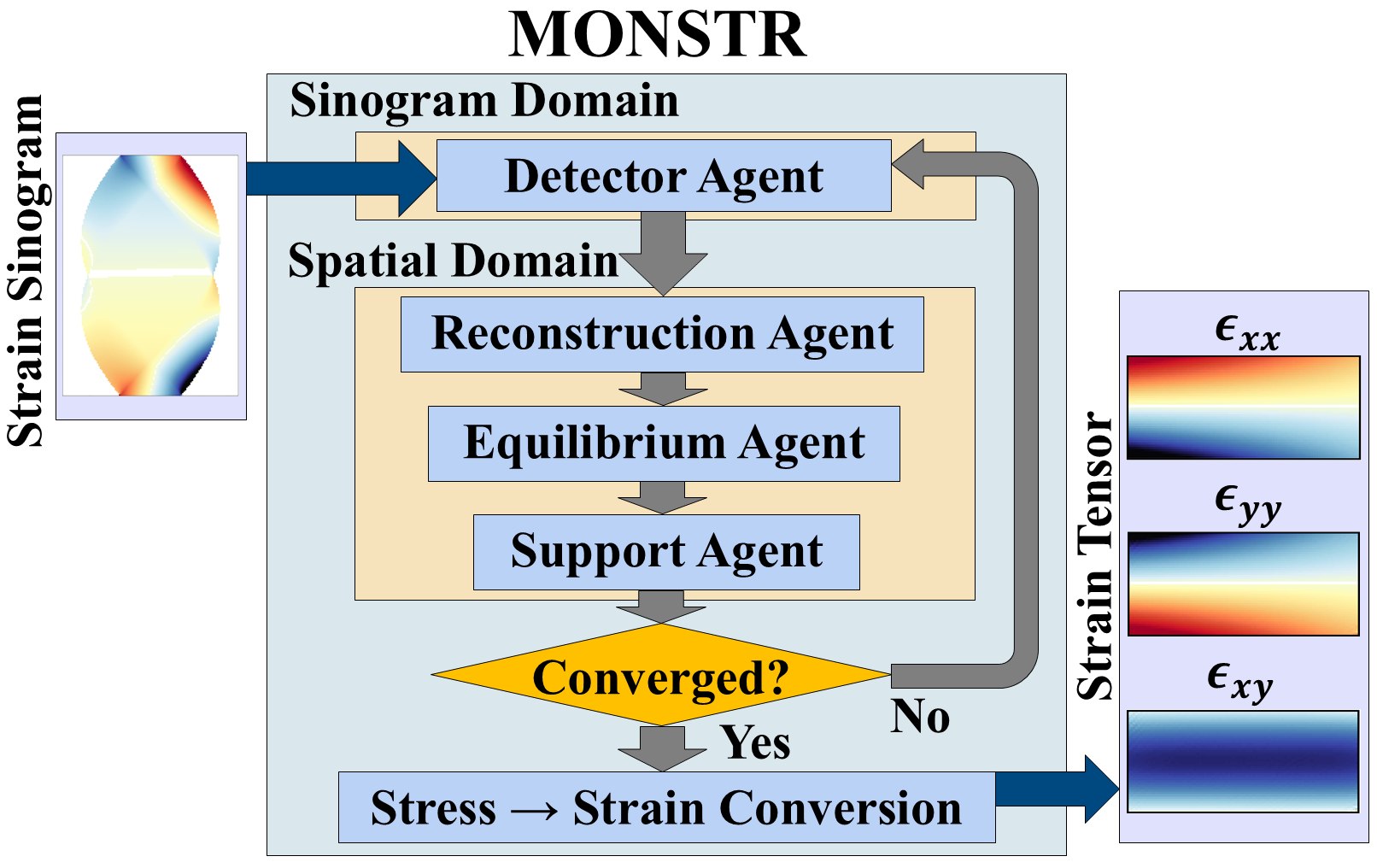}}
\vspace{-0.3cm}
\caption{MONSTR pipeline:
It uses a MACE framework that incorporates 4 agents to impose different constraints.
It iteratively executes these agents, trying to reach a consensus.}
\label{fig:monstr_pipeline}
\end{figure}

\section{ Model-Oriented Neutron Strain Tomographic Reconstruction (MONSTR)}
\label{sec:monstr}

While the forward model is described in terms of the strain tensor, we formulate our reconstruction algorithm, MONSTR, in terms of the stress tensor to simplify the mathematical description.
We define the following notation for stress:
\begin{itemize}[leftmargin=1em]
\vspace{-0.05cm}
\itemsep-0.4em
    \item \size{9.3} {$\sigma = [\sigma_{xx},\tab \sigma_{yy},\tab \sigma_{xy} ]^\top$ is the stress tensor. $\sigma_k$ is the scalar stress field along $k$, where $k=xx,yy,xy$.}
    \item \size{9.3} {$\tilde{p}_k = A\sigma_k$, where $k=xx,yy,xy$.}
    \item \size{9.3} {$\tilde{p} = [\tilde{p}_{xx},\tab \tilde{p}_{yy},\tab \tilde{p}_{xy} ]^\top$ is the tensor of stress virtual sinogram.}
\vspace{-0.05cm}
\end{itemize}
Now, under 2D plain-stress assumption, the stress-strain relationship at the $j^{th}$ spatial point is given by Hooke’s law \cite{sitharam2021theory}:
\begin{gather} \label{eq:strss_strain}
    \nonumber
     \begin{bmatrix} \sigma_{xx,j} \\ \sigma_{yy,j} \\ \sigma_{xy,j} \end{bmatrix} = \frac{E}{(1-\nu^2)} \begin{bmatrix} 1 & \nu & 0 \\ \nu & 1 & 0 \\ 0 & 0 & (1-\nu) \end{bmatrix}\begin{bmatrix} \epsilon_{xx,j} \\ \epsilon_{yy,j} \\ \epsilon_{xy,j} \end{bmatrix} \\
     \sigma_j = C \epsilon_j \ ,
\end{gather}
where $E$ is Young’s modulus, $\nu$ is Poisson’s ratio, and $C$ is the stiffness matrix.

MONSTR uses a MACE framework \cite{bouman2022mbir} to reconstruct the strain tensor.
MACE operates by coordinating multiple computational agents, each enforcing specialized constraints, to iteratively refine the solution until a consensus is reached.
For MONSTR, we used the following agents:
\begin{itemize}[leftmargin=1em]
\vspace{-0.05cm}
\itemsep-0.4em
    \item \textbf{Detector agent:} Obtains an estimate for the stress virtual sinogram tensor $\tilde{p}$ from the measurements $\langle\epsilon\rangle$.
    \item \textbf{Reconstruction agent:} Uses model-based iterative reconstruction (MBIR) \cite{bouman2022mbir} to reconstruct each component of the stress tensor $\sigma$ from the respective virtual sinograms $\tilde{p}$.
    \item \textbf{Equilibrium agent:} Encourages a soft equilibrium constraint on $\sigma$.
    \item \textbf{Support agent:} Sets the values of the stress tensor to zero in regions outside the sample. 
\vspace{-0.05cm}
\end{itemize}
MONSTR iteratively executes these agents until it reaches convergence, where the fixed point indicates consensus among all the agents.
However, as with any regularized inversion method, it is not guaranteed to reconstruct the exact strain/stress field.
Fig.~\ref{fig:monstr_pipeline} illustrates the entire process.

\subsection{Detector Agent}
The detector agent estimates the stress virtual sinogram tensor $\tilde{p}$ from the strain sinogram $\langle\epsilon\rangle$.
To implement it, we first need to express the forward model in terms of $\tilde{p}$.
First, we establish a relationship between $\tilde{p}$ and $p$ using the stiffness matrix defined earlier:
\begin{equation}
    \tilde{p_i} = C p_i \ .
\end{equation}
Using this relationship, we can now redefine the forward model from eq. \ref{eq:forward} in terms of $\tilde{p_i}$:
\begin{align} \label{eq:forward_stress}
    \bar{y}_i = \langle\epsilon\rangle_i L_i & = w_i p_i = (w_iC^{-1})(C p_i) = \tilde{w}_i \tilde{p}_i \ ,
\end{align}
where $\bar{y}_i$ is the ideal measurement and $\tilde{w}_i = w_iC^{-1}$ is the adjusted weight array for $\tilde{p_i}$.
The actual physical measurement is given by $y_i = \bar{y}_i + \text{noise}$.
For the $i^{th}$ sinogram entry, the detector agent solves the following optimization problem:
\begin{equation} \label{eq:det_prox_map}
    F_d(\tilde{p}_i^o) = \argmin_{\tilde{p}_i} \left \{ \frac{1}{2 \alpha_y^2}   \left( y_i - \tilde{w}_i\tilde{p}_i \right)^2 + \Vert  \tilde{p}_i - \tilde{p}_i^o \Vert^{2}_{2} \right \} \ ,
\end{equation}
where $\tilde{p}_i^o$ is some initial estimate, $\alpha_y$ is the factor controlling the impact of the data fitting term.
To solve this problem, we derived a point-wise closed-form solution and applied it to each sinogram entry.

\subsection{Reconstruction Agent}

The reconstruction agent transforms each component of the stress virtual sinogram tensor $\tilde{p}$ into each component of the spatial stress tensor $\sigma$.
We use MBIR for the tomographic reconstruction because it is capable of producing high-quality reconstructions even from sparse-view data \cite{bouman2022mbir}.
For each component index $k=xx, yy, xy$, MBIR solves the optimization problem given by
\begin{equation} \label{eq:recon}
    F_r(\tilde{p}_k) = \argmin_{\sigma_k} \left \{ \frac{1}{2 \alpha_v^2} \Vert \tilde{p}_k - A\sigma_k \Vert^{2}_{2} + g(\sigma_k) \right \} \ ,
\end{equation}
where $\alpha_v$ is the assumed noise standard deviation, and $g(\sigma_k)$ is the q-Generalized Gaussian Markov random field (qGGMRF) prior model \cite{bouman2022mbir}.
The software implementation was performed using SVMBIR, a robust and high-performance Python package for MBIR \cite{svmbir-code}.
Thus, each component of the stress tensor is reconstructed independently in this step. 

\begin{algorithm}[b!]
\SetAlgoLined
    \hspace{-0.3cm}\KwInput{$\langle\epsilon\rangle, L, \theta, \nu, m, \alpha_y, \alpha_v, \alpha_e$}
    \hspace{-0.3cm}\KwOutput{$\epsilon$}

    \hspace{-0.3cm}$y_i = \langle\epsilon\rangle_i L_i$ //Scale by path length 
    
    \hspace{-0.3cm}$w = \mbox{compute\_weights}[\theta]$ //Forward model weights
    
    \hspace{-0.3cm}$C = \mbox{comp\_stiffness\_matrix}[\nu]$ //To convert strain-stress
    
    \hspace{-0.3cm}$\tilde{w}_i = w_i C^{-1}$

    \hspace{-0.3cm}Init: $\sigma, u$
    
    \vspace{0.1cm}
    \Repeat{
   convergence = True
    }
    {\tcp{Sinogram domain constraints}
    
    $\tilde{p} \gets F_d(\mbox{Proj}(\sigma)-u; y, \tilde{w}, \alpha_y)$ //Data fitting
        
    \vspace{0.15cm}
    \tcp{Spatial domain constraints}
        
    $\sigma \gets F_r(\tilde{p} + u; \alpha_v)$ //Tomographic reconstruction
        
    $\sigma \gets F_e(\sigma; \alpha_e)$ //Promote equilibrium constraint 
        
    $\sigma \gets F_s(\sigma; m)$ //Ensure zero outside the sample 
    
    \vspace{0.1cm}
    $u \gets u + \tilde{p} - \mbox{Proj}(\sigma)$}
    
    \vspace{0.1cm}
    \hspace{-0.3cm}\Return{$\epsilon = C^{-1}\sigma$}
\caption{MONSTR}
\label{alg:monstr_detailed}
\end{algorithm}

\subsection{Equilibrium Agent}
The equilibrium agent promotes a soft equilibrium constraint on $\sigma$ while staying close to some initial estimate.
It is known from continuum mechanics that stress fields should adhere to the equilibrium constraint at each location in the sample. 
The 2D stress equilibrium constraint for continuous case \cite{sitharam2021theory} can be expressed as:
\begin{align}
    \label{eq:equilibrium_eq_1}
    \frac{\partial}{\partial x} \sigma_{xx}+ \frac{\partial}{\partial y} \sigma_{xy} = 0\\
    \label{eq:equilibrium_eq_2}
    \frac{\partial}{\partial y} \sigma_{yy}+ \frac{\partial}{\partial x} \sigma_{xy} = 0
\end{align}
We note that the same notation has been used for both continuous and discrete stress tensor fields to avoid introducing new variables.
We can represent \ref{eq:equilibrium_eq_1} and \ref{eq:equilibrium_eq_2} using a single equation as:
\begin{gather} \label{eq:equilibrium_op}
    \nonumber
    \begin{bmatrix} \frac{\partial}{\partial x} & 0 & \frac{\partial}{\partial y} \\ 0 & \frac{\partial}{\partial y} & \frac{\partial}{\partial x} \end{bmatrix}\ \begin{bmatrix} \sigma_{xx} \\ \sigma_{yy} \\ \sigma_{xy}  \end{bmatrix} =0 \\
    \mathbb{D}(\sigma) = 0 \ ,
\end{gather}
where $\mathbb{D}( )$ is the equilibrium operator.
Now, we can express the equilibrium agent as:
\begin{equation} \label{eq:equi_proj_op}
    F_e(\sigma^o) = \argmin_{\sigma} \left \{ \frac{1}{2 \alpha_e^2} \Vert \mathbb{D}(\sigma) \Vert^{2}_{2} + \Vert \sigma - \sigma^o \Vert^{2}_{2} \right \} \ ,
\end{equation}
where $\sigma^o$ is the initial estimate and 
$\alpha_e$ controls the relative strength between the two terms.
Since it is difficult to derive a direct closed-form solution to this problem, we decompose the original optimization problem into three smaller subproblems, each corresponding to a single stress component. 
MONSTR applies closed-form solutions to these subproblems within a coordinate descent framework to update the stress tensor.

\subsection{Support Agent}
The support agent imposes a shape constraint on the reconstructed stress tensor $\sigma$ by setting values outside the sample to zero using a binary shape mask.
This mask is typically assumed to be known \cite{gregg2018general, hendriks2019general} and can be obtained by performing a standard CT reconstruction of the sample, followed by thresholding.
For the $k^{th}$ stress component at $j^{th}$ spatial point, the operator performs the following operation:
\begin{equation}
    F_s(\sigma_{k,j}^o) = m_j \sigma_{k,j}^o \ ,
\end{equation}
where $m_j$ is the known binary shape mask value at the $j^{th}$ point.
Algorithm \ref{alg:monstr_detailed} shows a detailed representation of MONSTR, where $\mbox{Proj}(\sigma)=[A\sigma_{xx},\tab A\sigma_{yy},\tab A\sigma_{xy} ]^\top$.

\section{Results}
\label{sec:results}

\begin{figure}[b!]
\vspace{-0.4cm}
\centering
\includegraphics[width=.65\linewidth]{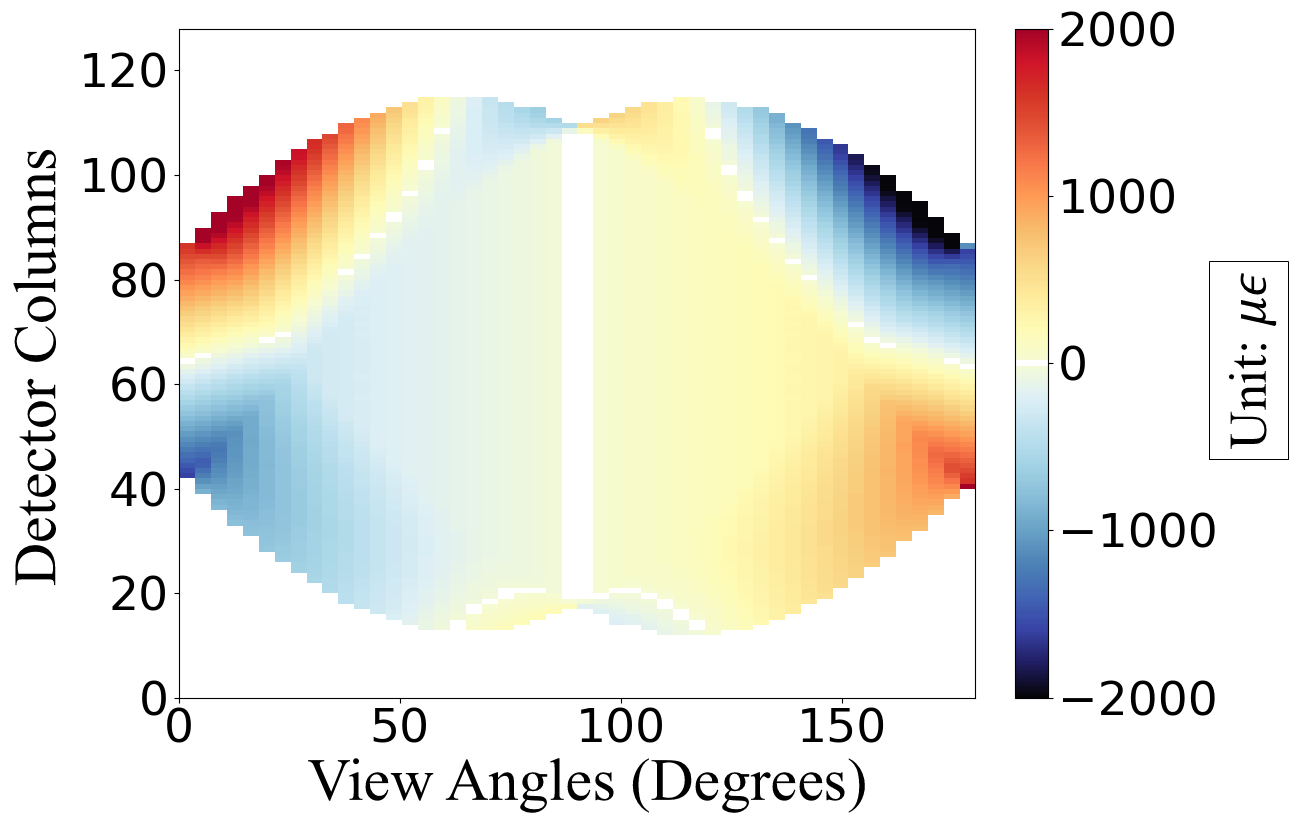}
\vspace{-0.15in}
\caption{Simulated strain sinogram (50 views, 128 columns).}
\label{fig:strain_sino}
\end{figure}

\begin{figure*}[t!]
\centering
\centerline{\includegraphics[width=0.95\linewidth]{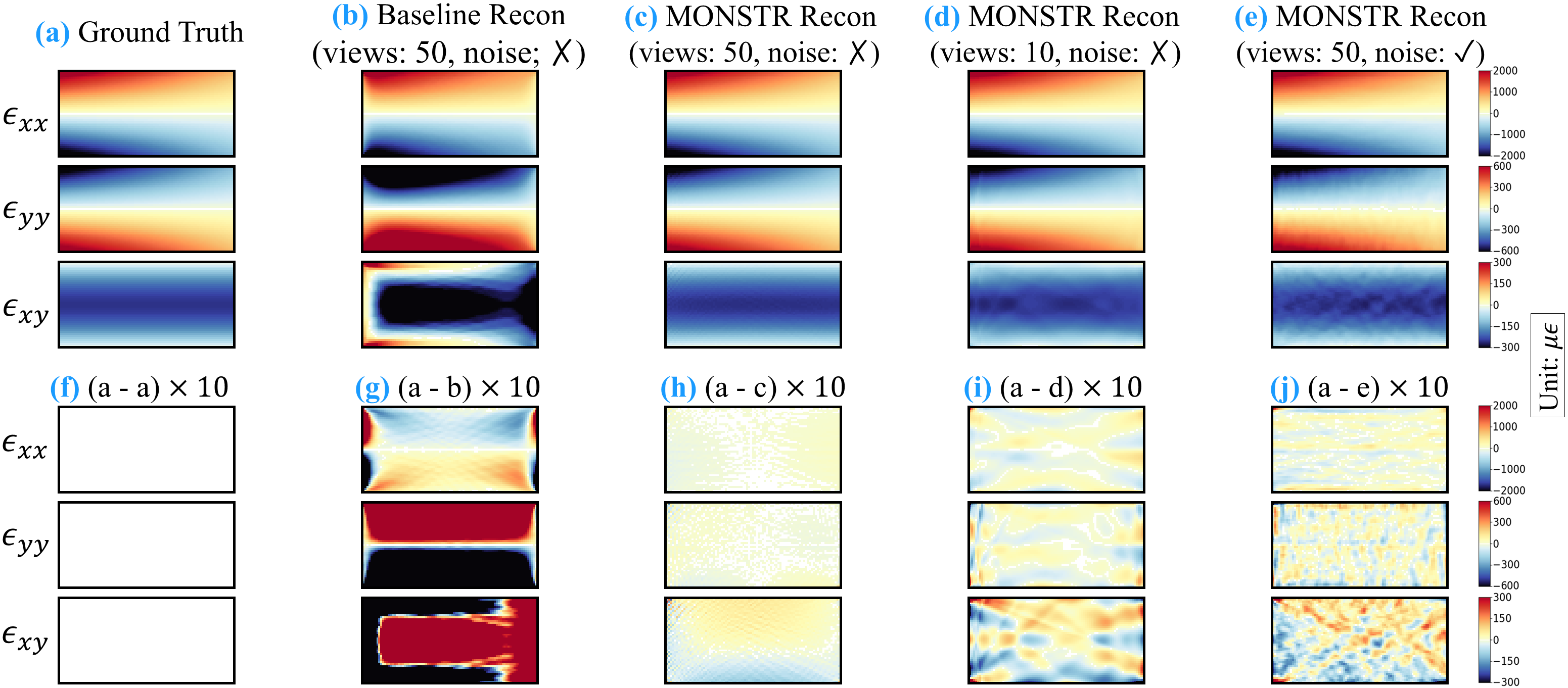}}
\vspace{-0.4cm}
\caption{Strain fields: (a) simulated ground truth; (b) baseline reconstructions from noiseless sinogram with 50 views; (c)–(e) MONSTR reconstructions from different input sinograms: (c) noiseless with 50 views, (d) noiseless with 10 views, and (e) noisy with 50 views.
(f)–(j) show the corresponding error images compared to the ground truth, scaled by 10.
MONSTR clearly outperforms the baseline and remains effective even for input sinograms with fewer views or added noise.}
\vspace{-0.15in}
\label{fig:recon_strain}
\end{figure*}

We used simulated data for validation purposes.
The simulation process is based on the classical cantilevered beam described in \cite{gregg2018general}.
Under a plane-stress assumption, the Saint-Venant approximation to the simulated strain tensor \cite{beer1992simulation} at the $j^{th}$ point is:
\begin{equation} \label{eq:simulation}
    \nonumber
    \begin{bmatrix} \epsilon_{xx,j} \vspace{0.07cm} \\ \epsilon_{yy,j} \vspace{0.07cm} \\ \epsilon_{xy,j} \end{bmatrix} = \begin{bmatrix} \frac{P}{EI}(l-x_j)y_j \vspace{0.07cm}\\ -\frac{(1+\nu)P}{2EI}[(\frac{h}{2})^2-y_j^2] \vspace{0.07cm}\\ -\frac{\nu P}{EI}(l-x_j)y_j  \end{bmatrix} \ ,
\end{equation}
where $P$ is the applied load, $I$ is the second moment of area, $l$ is the sample length, and $h$ is the sample width.
$(x_j, y_j)$ is the 2D coordinate of the $j^{th}$ point.
The overall field of view has a size of $128 \times 128$ voxels, and the actual sample only occupies $45 \times 91$ voxels within this field of view. 
Finally, using the forward model presented in eq. \ref{eq:forward}, we compute a strain sinogram with 50 projection views (between 0 and 180 degrees) and 128 detector columns as shown in Fig.~\ref{fig:strain_sino}.

We applied MONSTR to the simulated strain sinogram and reconstructed the strain tensor.
Since prior methods \cite{gregg2018general, hendriks2019general} lack sufficient implementation details and open-source code, direct comparisons were not feasible.
As an alternative, we implemented a simple baseline model that excludes the equilibrium constraint and compared its performance with MONSTR.
Additionally, to evaluate MONSTR’s effectiveness in reducing the number of required views—and thereby scan time—we also tested it on a strain sinogram subsampled to only 10 views.
In order to assess robustness to noise, we also conducted an experiment by adding white Gaussian noise with a standard deviation of $10\mu\epsilon$ to the 50-view sinogram and applied MONSTR to the noisy data.

Fig.~\ref{fig:recon_strain} shows the ground truth strain components and the corresponding reconstructions for all cases.
It also includes the $10\times$-scaled error images relative to the ground truth for each case.
Comparing the error images in (g) and (h), it is evident that MONSTR reconstructions in (c) are significantly more accurate than the baseline reconstructions in (b), closely matching the simulated ground truth in (a).
Furthermore, MONSTR reconstructions using input sinograms with fewer views (d) or added noise (e) still demonstrate reasonably good quality with minimal artifacts.
This robustness can be attributed to the use of MBIR, which is well-suited for reconstructing from noisy or sparse-view data.

\begin{figure}[b!]
\vspace{-0.2in}
\centering
\centerline{\includegraphics[width=0.8\linewidth]{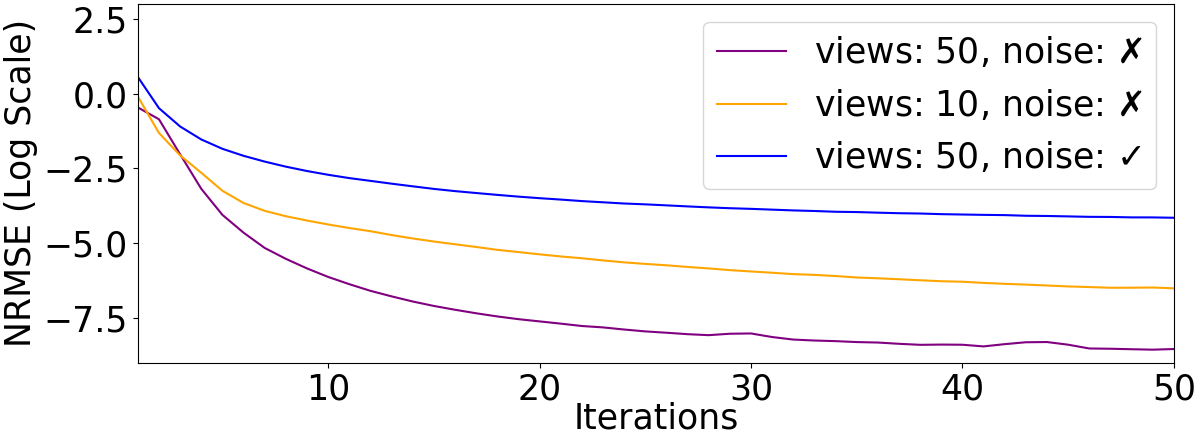}}
\vspace{-0.15in}
\caption{Convergence of MONSTR: By plotting the NRMSE between $\tilde{p}$ and $\mbox{Proj}(\sigma)$ across iterations, we observe satisfactory convergence for all cases.}
\label{fig:convergence}
\end{figure}

\begin{table}[t!]
\begin{center}
\caption{NRMSE between ground truth and reconstructions.\\}
\label{table:nrmse}
\begin{tabular}{|@{}c@{}|c|c|c|c|}
  \hline
  NRMSE & $\epsilon_{xx}$ & $\epsilon_{yy}$ & $\epsilon_{xy}$ & $\epsilon$\\
  \hline
  baseline & \multirow{2}{*}{0.1359} & \multirow{2}{*}{0.4557} & \multirow{2}{*}{0.4815} & \multirow{2}{*}{0.2612}\\[-4pt]
  \size{9} { (views: 50, noise: \ding{55}) } &&&&\\
  \hline
  MONSTR & \multirow{2}{*}{\textbf{0.0055}} & \multirow{2}{*}{\textbf{0.0079}} & \multirow{2}{*}{\textbf{0.0233}} & \multirow{2}{*}{\textbf{0.0072}}\\[-4pt]
  \size{9} { (views: 50, noise: \ding{55}) } &&&&\\
  \hline
  MONSTR & \multirow{2}{*}{0.0239} & \multirow{2}{*}{0.0365} & \multirow{2}{*}{0.0565} & \multirow{2}{*}{0.0269}\\[-4pt]
  \size{9} { (views: 10, noise: \ding{55}) } &&&&\\
  \hline
  MONSTR & \multirow{2}{*}{0.0213} & \multirow{2}{*}{0.0415} & \multirow{2}{*}{0.0531} & \multirow{2}{*}{0.0251}\\[-4pt]
  \size{9} { (views: 50, noise: \ding{51}) } &&&&\\
  \hline
\end{tabular}
\end{center}
\vspace{-0.5cm}
\end{table}

Table~\ref{table:nrmse} summarizes the quantitative performance across different cases.
For the noiseless 50-view case, MONSTR achieves a normalized root mean square error (NRMSE) of 0.0072—substantially lower than the baseline's NRMSE of 0.2612.
In the reduced-view and noisy scenarios, the NRMSE increases to 0.0269 and 0.0251, respectively.
While these values are higher, they remain sufficiently low to demonstrate MONSTR’s robustness to both view sparsity and noise.

We ran MONSTR for 50 iterations in all cases.
Each iteration took approximately 0.94 seconds on an Apple device with an M2 Max chip (12-core CPU, 38-core GPU) and 64 GB of memory.
For each iteration, we computed the NRMSE between $\tilde{p}$ and $\mbox{Proj}(\sigma)$ (the terms used to update $u$ in the last step of each outer iteration in Alg. 1) as an indicator of convergence.
Here, $\tilde{p}$ is derived from the sinogram domain agent, while $\mbox{Proj}(\sigma)$ originates from the spatial domain agents.
So, a decreasing NRMSE indicates growing consensus among the agents.
As shown in Fig.~\ref{fig:convergence}, the NRMSE consistently decreases and reaches convergence within 50 iterations in all cases.
However, the final NRMSE is lowest for the noiseless 50-view case and highest for the noisy 50-view case.

\vspace{-0.1in}
\section{Conclusion}
\vspace{-0.1in}
\label{sec: conclusion}
We introduce MONSTR, an algorithm for reconstructing 2D residual strain tensor using Bragg edge imaging.
The algorithm modularizes the problem within a MACE framework, leveraging MBIR for high-quality reconstructions.
Additionally, it incorporated equilibrium and support constraints to reduce the degrees of freedom in the inverse problem.
When applied to simulated data, MONSTR produced high-quality reconstructions of the strain components with an overall NRMSE of less than 1\% compared to the ground truths.

\section{Acknowledgment}
\vspace{-0.1in}
C. Bouman was partially supported by the Showalter Trust. 
This research used resources at the Spallation Neutron Source, a DOE Office of Science User Facility operated by the Oak Ridge National Laboratory.

\bibliographystyle{IEEEbib}
\bibliography{icip_2025}
\end{document}